\documentclass{webofc}
\usepackage[varg]{txfonts}   
\begin{document}
\vspace*{15mm}
\hspace{8.3cm}{\small{\textrm{YITP-22-107, J-PARC-TH-0277}}}
\vspace*{-13mm}

\title{From lattice to observables: Real and virtual experiments for exploring hot and dense QCD}


\author{\firstname{Masakiyo} \lastname{Kitazawa}\inst{1,2,3}\fnsep\thanks{\email{kitazawa@yukawa.kyoto-u.ac.jp}}
}

\institute{
  Yukawa Institute for Theoretical Physics, Kyoto University, 
  606-8502, Kyoto, Japan
  \and
  J-PARC Branch, KEK Theory Center, Institute of Particle and Nuclear Studies, KEK, 203-1,
  319-1106, Shirakata, Tokai, Ibaraki, Japan
  \and
  Department of Physics, Osaka University, 
  560-0043, Toyonaka, Osaka, Japan
}

\abstract{%
  Relativistic heavy-ion collisions and the lattice QCD Monte-Carlo simulations
  are important ``experimental'' tools for investigating the properties of
  the medium described by QCD under extreme conditions.
  After briefly examining their characteristics, I pick up 
  the study of the critical points in QCD as an example of the research
  subjects that these two tools play complementary roles and discuss
  recent topics that have been achieved with the use of each tool.
}
\maketitle
\section{Introduction}
\label{intro}

Exploring strongly-interacting medium under extreme conditions is one of the main topics of Quantum Chromodynamics (QCD) and hadron physics. The hadronic medium undergoes a phase transition to a new form of matter called the quark-gluon plasma (QGP) or the quark matter in which the quarks and gluons are liberated from the confinement into hadrons and act as basic degrees of freedom of the system. This phase transition is expected to occur when the temperature $T$ or the baryon-number density $\rho$ exceeds $T\simeq150$~MeV~$\simeq1.5\times10^{12}$~K or $\rho\simeq10^{15}$~g/cm$^3$~\cite{Yagi:2005yb}. Understanding their properties is indispensable for various purposes, such as investigations of the form of the matter in the early Universe and inside neutron stars.

Although these conditions are far from the environments around us, we have experimental tools to investigate these extreme conditions. Roughly speaking, they are classified into two types. One is the relativistic heavy-ion collisions (HIC) that collide two heavy nuclei accelerated up to almost the speed of light. These experiments have been performed and are ongoing all over the world at the Large Hadron Collider (LHC), Relativistic Heavy Ion Collider (RHIC), etc. Other experiments are the first-principle numerical simulations based on the lattice QCD, which are virtual experiments on supercomputers. Nowadays, numerical simulations with physical quark masses and their extrapolations to the continuum limit are possible for various observables. They thus are regarded as the ``experimental'' results. In the following, experiments in this category are denoted as the LAT.

For exploring the hot and dense medium, the complementary use of these two experiments is highly effective, because the HIC and LAT have different characteristics, advantages and disadvantages. Now, let us discuss their properties and pros/cons in some more detail using Table~\ref{tab-1} that shows a short summary.

First, needless to say, the HIC are real experiments, which is a clear advantage compared with the latter. On the other hand, the fact that the LAT are simulations can become their unique advantage because the LAT can deal with experiments of the unphysical world, such as QCD with unphysical quark masses; such a study sometimes gives us novel insights into our world.

\begin{table}
\centering
\caption{Short summary of the charcteristics of the HIC and LAT.}
\label{tab-1}       
\begin{tabular}{r|l}
\hline
HIC & LAT \\
\hline
\hline
real & virtual \\
dynamical & thermal \\
\hline
\multicolumn{2}{c}{~~~$\mu_B\ne0$}\\
possible & difficult \\
\hline
\end{tabular}
\end{table}

Second, the HIC produce highly-dynamical systems that expand rapidly after the creation. To extract physical information on the hot medium that exists only in the early stage of this expansion, therefore, a description of the whole dynamics of the space-time evolution is inevitable. In contrast, the LAT perform simulations of completely equilibrium systems. This allows the LAT to investigate thermal properties of the medium, such as the equation of states~\cite{Borsanyi:2013bia}, in an ideal environment. These results are used as useful inputs for the dynamical modeling of the HIC. However, this fact can also be a disadvantage of the LAT, because it is difficult for the LAT to deal with non-equilibrium systems. For example, the quantitative analysis of the transport coefficients is still a challenging subject in the LAT, although they are important observables in describing the dynamics of the HIC.

Finally, the HIC are capable of investigating various $T$ and $\rho$, since these parameters can be varied by changing the collision energy; for $\sqrt{s_{_{NN}}}\gtrsim3$~GeV, the baryon-number density of the medium produced by the collisions monotonically increases by reducing $\sqrt{s_{_{NN}}}$. Series of the HIC with various collision energies is called the beam-energy scan, and active experimental activities are ongoing for exploring various regions on the $T$--$\rho$ plane. For example, these experiments are ongoing at RHIC, NA61/SHINE and HADES, and they will be carried over to future experiments at FAIR and J-PARC-HI~\cite{Galatyuk:2019lcf}. On the contrary, the conventional algorithm of the LAT is applicable only for the system at zero baryon chemical potential $\mu_B$ because of the so-called sign (complex-phase) problem. Therefore, investigation of the system with large $\rho$ is difficult in the LAT.

\section{QCD Critical points}

In the following part of this manuscript, as an example of the research subjects where the HIC and LAT play useful roles in a complementary way, I would like to report on recent progress in the analysis of the critical points in QCD. Since the same mathematical concept for the probability distribution is used in these studies, their simultaneous explanation will also be useful for the mutual understanding.

\begin{figure}[t]
\centering
\includegraphics[width=.45\textwidth,clip]{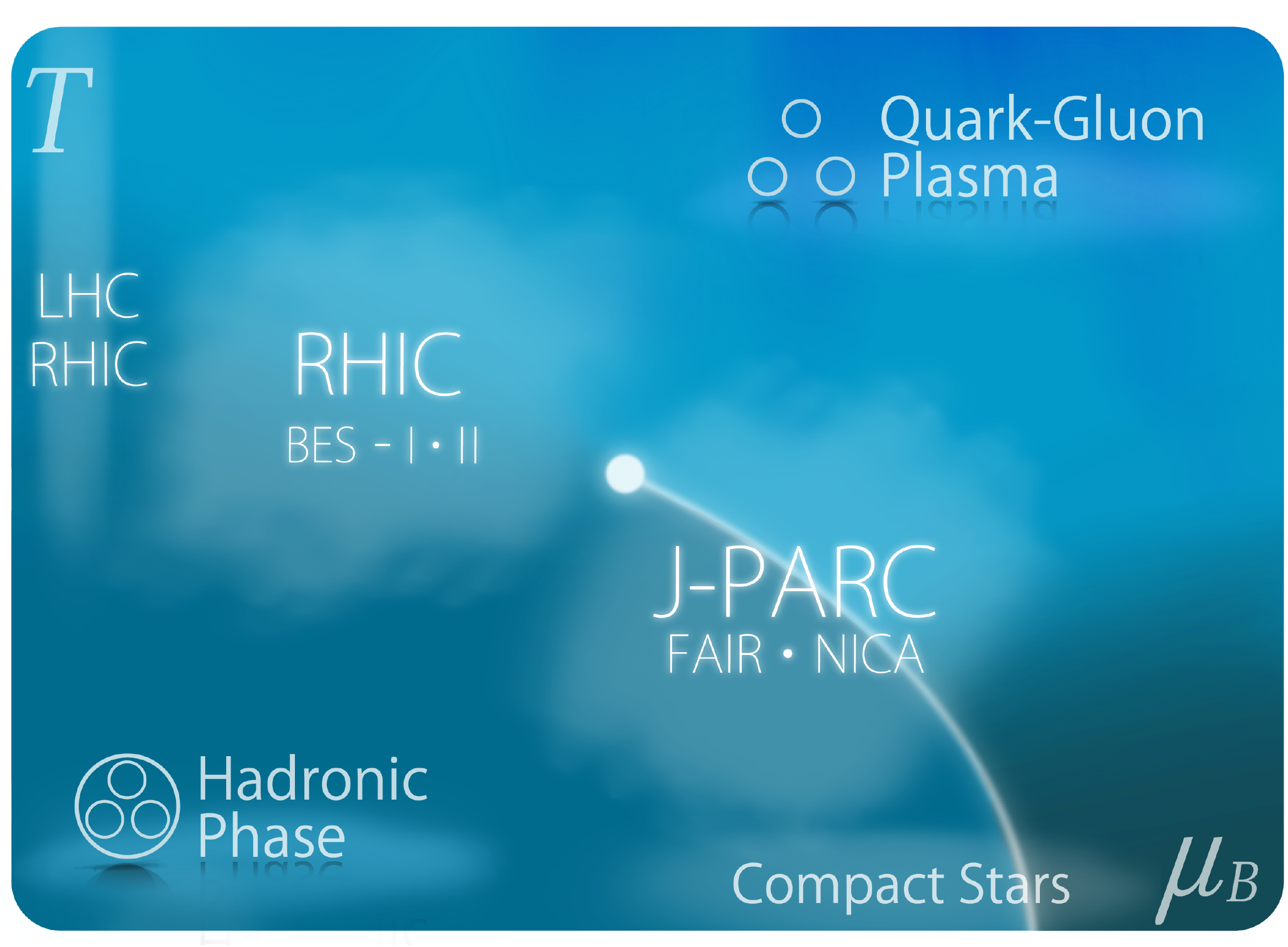}
\caption{Conceptual phase diagram of QCD on the $T$--$\mu_B$ plane~\cite{J-PARC-HI}.}
\label{fig:phased}
\end{figure}

An end-point of the first-order transition line to crossover is called the critical point (CP). The CP is a second-order transition point and singularities of various observables manifest themselves there. In the QCD phase diagram on the $T$--$\mu_B$ plane, the existence of the first-order transition line is expected at the large $\mu_B$ region as shown in Fig.~\ref{fig:phased}. If this is the case, the CP does exist and it is called the QCD CP. The experimental search for the CP is one of the central subjects of the beam-energy scan in the HIC~\cite{Asakawa:2015ybt,Bluhm:2020mpc}.

The appearance of CPs in QCD is not limited to the one shown in Fig.~\ref{fig:phased}. Besides the possible existence of other CP on the $T$--$\mu_B$ plane~\cite{Kitazawa:2002jop}, CPs can appear with variations of the quark masses and other external parameters. For example, whereas the phase transition of QCD at $\mu_B=0$ is crossover at the physical quark masses, the phase transition becomes first order in the massless three-flavor QCD and the heavy-mass limit. This means that CPs do exist in the phase diagram of quark masses~\cite{Gavin:1993yk}. Their attractive feature is that they exist at $\mu_B=0$, and hence their investigation in the LAT is feasible with the conventional method. For this reason, these critical points have been actively investigated by the LAT over the past several decades~\cite{Gavin:1993yk,Ejiri:2019csa,Kuramashi:2020meg,Kiyohara:2021smr}.

\section{Cumulants}
\label{sec:cum}

Since a CP is a second-order transition point, fluctuations of observables having a coupling with the order parameter show a divergence. Fluctuations are described by the probability distribution function, and the cumulants are useful quantities to denote its property~\cite{Asakawa:2015ybt}. In this section, we give a brief review of the cumulants as an introduction to the later sections.

Let us consider a real variable $x$ and its probability distribution function $P(x)$. A set of standard quantities to characterize $P(x)$ is the moments
\begin{align}
  \langle x^n \rangle = \int dx x^n P(x)
  = \partial_\theta^n \langle e^{x\theta} \rangle|_{\theta=0}.
\end{align}
Cumulants are another useful set of quantities to describe $P(x)$ that are defined by
\begin{align}
  \langle x^n \rangle_{\rm c}
  = \partial_\theta^n \ln \langle e^{x\theta} \rangle|_{\theta=0}.
\end{align}
For lower orders, they are related to each other as
\begin{align}
  \langle x \rangle_{\rm c} = \langle x \rangle,
  \quad
  \langle x^2 \rangle_{\rm c} = \langle (\delta x)^2 \rangle,
  \quad
  \langle x^3 \rangle_{\rm c} = \langle (\delta x)^3 \rangle,
  \quad
  \langle x^4 \rangle_{\rm c} = \langle (\delta x)^4 \rangle - 3 \langle (\delta x)^2 \rangle^2,
\end{align}

Compared with the moments, the cumulants have various useful properties~\cite{Asakawa:2015ybt}. For example, cumulants in thermal systems are extensive variables, whereas the moments are not. For the Gauss distribution, the cumulants at the third and higher orders vanish; therefore, nonzero cumulants higher than the second order represent non-Gaussianity of $P(x)$. The third- and fourth-order cumulants are related to skewness $S$ and kurtosis $\kappa$ as 
\begin{align}
  S = \frac{\langle x^3 \rangle_{\rm c}}{\langle x^2 \rangle_{\rm c}^{3/2}},
  \qquad
  \kappa = \frac{\langle x^4 \rangle_{\rm c}}{\langle x^2 \rangle_{\rm c}^2}.
\end{align}
Also, for the Poisson distribution, all the cumulants are equivalent; $\langle x^n \rangle_{\rm c} = \langle x \rangle_{\rm c}$.

In Sec.~\ref{sec:LAT}, we use the quantity called the (fourth-order) Binder cumulant $B_4$, which is defined by
\begin{align}
  B_4 = \kappa + 3.
  \label{eq:B4}
\end{align}
For the Gauss distribution, one has $B_4=3$.

\section{Search for the QCD CP in the HIC}
\label{sec:HIC}

One of the most important subjects of the beam-energy scan in the HIC is the search for the QCD CP on the $T$--$\mu_B$ plane. Fluctuations of particle numbers are observable by the event-by-event analysis in the HIC, and it is believed that an enhancement of the fluctuations associated with the QCD CP is observable in these experiments. Besides the enhancement of the second-order cumulant, it is pointed out that yet higher order ones have more characteristic behavior around the CP, and thus are clearer signals of the CP~\cite{Stephanov:2008qz,Asakawa:2009aj}. Active experimental analyses on the cumulants have been performed recently~\cite{STAR:2019ans,Arslandok:2020mda}.

Among the fluctuation observables, those of conserved charges have particularly useful properties. First, the cumulants of conserved charges, such as the baryon number $N_{\rm B}$ and electric charge $N_{\rm Q}$, are well-defined and calculable unambiguously in QCD, whereas those of non-conserved quantities are not. This also means that the conserved-charge cumulants are observable on the LAT~\cite{Bellwied:2019pxh,Bollweg:2021vqf}. This allows one to make a direct comparison between the LAT and HIC in terms of the conserved-charge cumulants. Another advantage of the conserved-charge fluctuations is that their time evolution is typically slow, and hence the fluctuations developed near the CP can survive until the final state~\cite{Asakawa:2015ybt,Kitazawa:2013bta}.

\begin{figure}[t]
\centering
\includegraphics[width=.7\textwidth,clip]{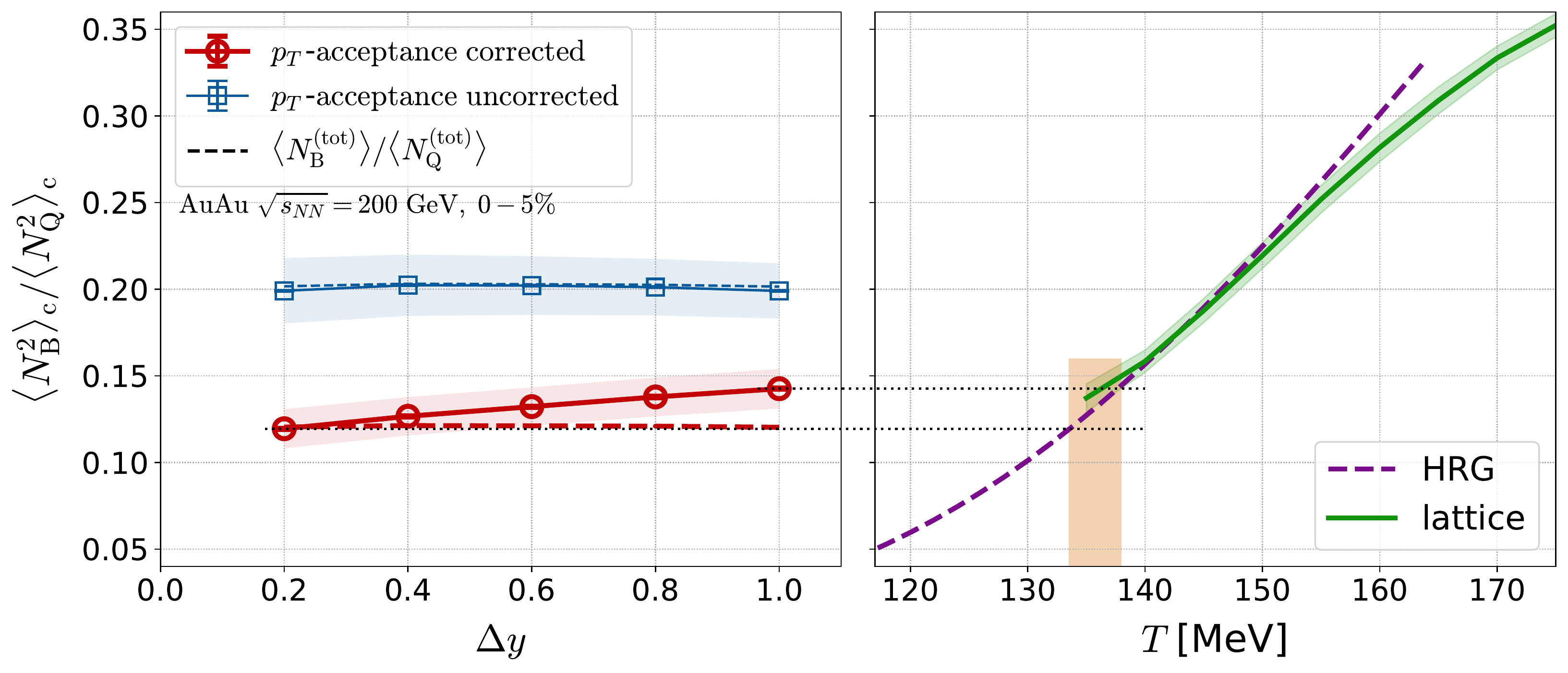}
\caption{
  Ratio of the second-order cumulants of the baryon number and electric charge
  $\langle N_{\rm B}^2 \rangle_{\rm c} / \langle N_{\rm Q}^2 \rangle_{\rm c}$
  constructed from the experimental data by the STAR Collaboration
  at $\sqrt{s_{_{NN}}}=200$~GeV and its comparison with the lattice results~\cite{Kitazawa:2022gmq}.}
\label{fig:chiBQ}
\end{figure}

In the HIC, however, a direct measurement of the conserved-charge cumulants is difficult. In particular, the measurement of the baryon number $N_{\rm B}$ is problematic since typical detectors in the HIC cannot observe neutrons that are neutral. For this reason, the proton number cumulants are measured and compared with theoretical results on the baryon number cumulants. As pointed out in Ref.~\cite{Kitazawa:2011wh}, however, they are different quantities. Similarly, imperfect efficiencies and acceptance of experimental detectors affect the measurement. Although the effects of the former are usually corrected~\cite{Nonaka:2017kko}, the latter's effect has not been considered seriously in the literature.

To make a solid comparison between the LAT and HIC, in Ref.~\cite{Kitazawa:2022gmq} we constructed the ratio of the second-order cumulants of the baryon number and electric charge $\langle N_{\rm B}^2 \rangle_{\rm c} / \langle N_{\rm Q}^2 \rangle_{\rm c}$ from the experimental data of STAR Collaboration for $\sqrt{s_{_{NN}}}=200$~GeV~\cite{STAR:2019ans} by correcting the effect of the acceptance cut on the transverse momentum. The corrected result for $\langle N_{\rm B}^2 \rangle_{\rm c} / \langle N_{\rm Q}^2 \rangle_{\rm c}$ is shown in the left panel of Fig.~\ref{fig:chiBQ} by the red circles~\cite{Kitazawa:2022gmq}. By assuming that the cumulant ratio reflects a single temperature, the result can be compared with the result obtained on the LAT~\cite{Bollweg:2021vqf}. This comparison is made in Fig.~\ref{fig:chiBQ}, and it shows that the value of $T$ extracted from the comparison is $T\simeq134-138$~MeV. Although it is sometimes assumed that the fluctuations observed in the HIC are generated at the chemical freezeout temperature $T_{\rm chem}\simeq155$~MeV, Fig.~\ref{fig:chiBQ} shows that this assumption is not justified.

\section{Study of QCD critical point in LAT}
\label{sec:LAT}

\begin{figure}[t]
\centering
\includegraphics[width=.4\textwidth,clip]{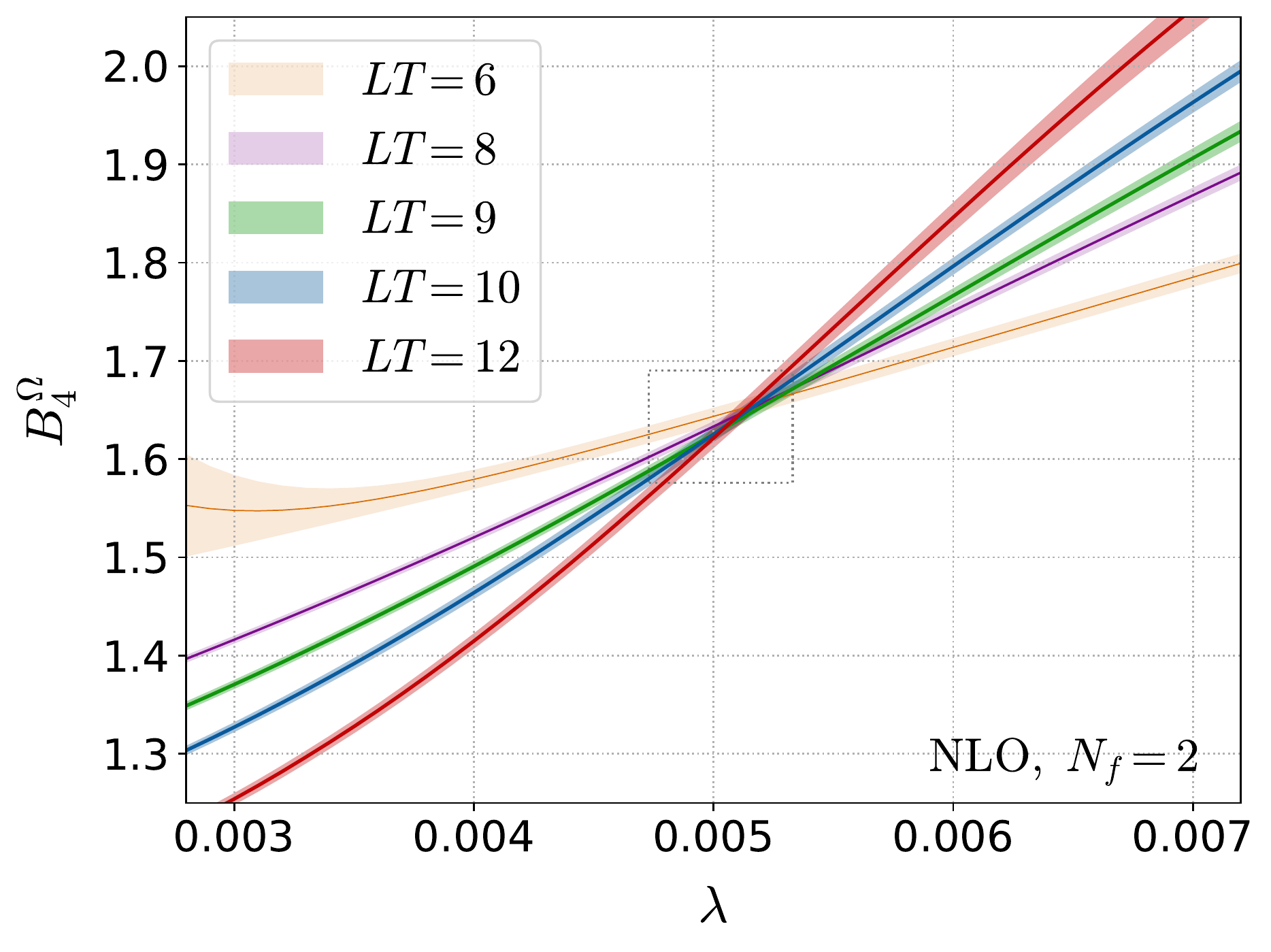}
\includegraphics[width=.4\textwidth,clip]{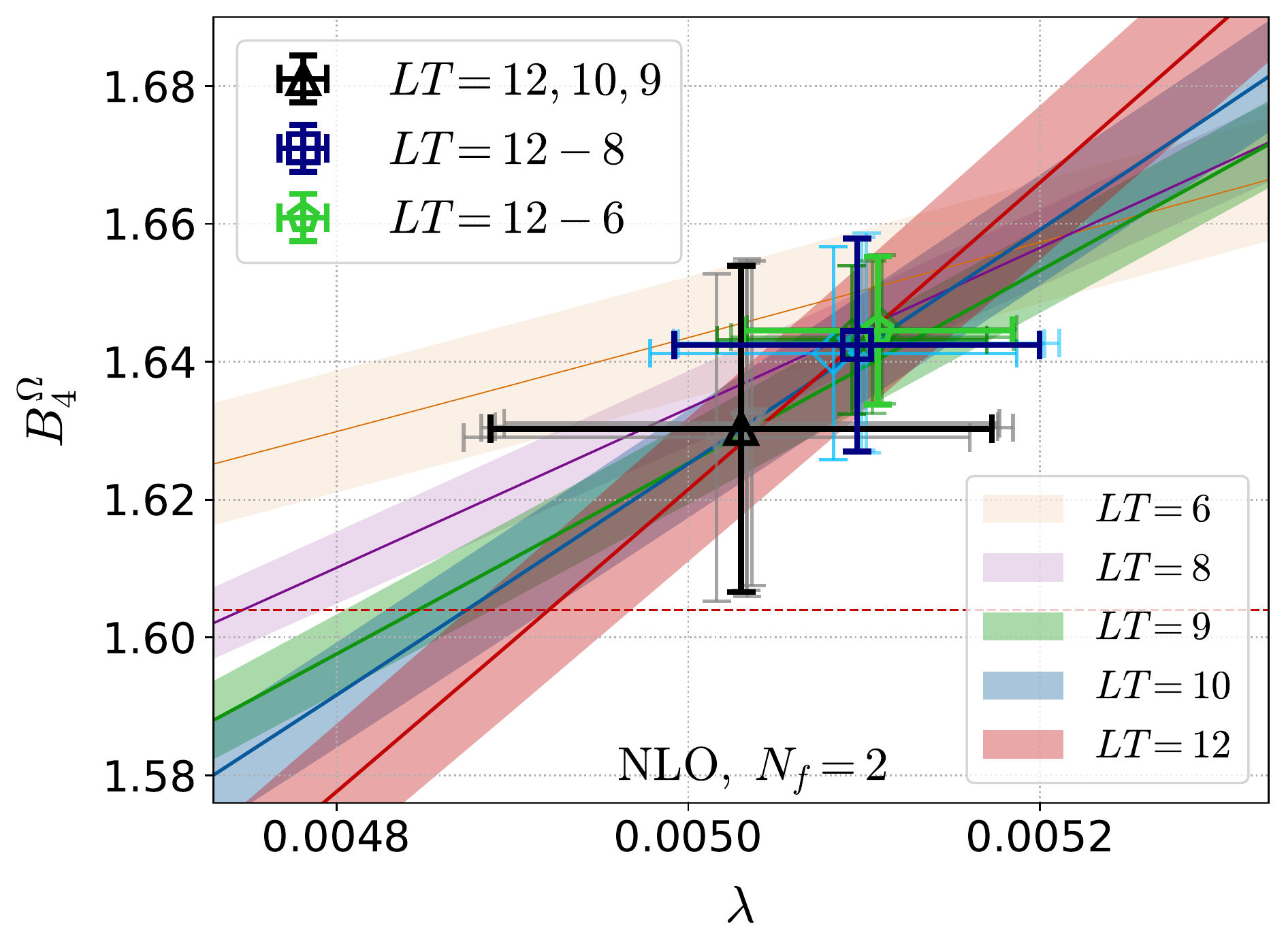}
\caption{Finite-size scaling analysis of the Binder cumulant $B_4$~\cite{Kiyohara:2021smr}.}
\label{fig:B4}       
\end{figure}

The direct analysis of the CP on the $T$--$\mu_B$ plane in the LAT is difficult due to the sign problem. On the other hand, the CPs that appear with the variations of the quark masses at $\mu_B=0$ can be investigated in detail with the conventional algorithms. For this reason, these CPs and the phase structure on the quark-mass plane have been investigated in the LAT over the past decades~\cite{Gavin:1993yk,Ejiri:2019csa,Kuramashi:2020meg,Kiyohara:2021smr}. 

One of the conventional methods to study the critical point in numerical simulations is the Binder cumulant analysis. In this method, one focuses on the fourth-order Binder cumulant $B_4$ defined in Eq.~(\ref{eq:B4}). On the transition line, this quantity behaves as
\begin{align}
  B_4 = 1 ~~\mathrm{(first~order~side)},
  \qquad
  B_4 = 3 ~~\mathrm{(crossover~side)},
  \label{eq:B4:13}
\end{align}
in the thermodynamic limit. Therefore, the discontinuous change of the value of $B_4$ can be used for locating the critical point. However, on the finite volume the relation~(\ref{eq:B4:13}) is violated and the discontinuity is smoothen. Even in this case, from the finite-size scaling (FSS) argument it is known that $B_4$ of the magnetic observable behaves as
\begin{align}
  B_4(t) = b_4 + c(t-t_c)L^{1/\nu} + {\cal O}((t-t_c)^2),
  \label{eq:B4t}
\end{align}
where $t$ is a parameter along the critical line and $t_c$ represents the location of the CP, $L$ is the spatial extent of the system, and $b_4$ and $\nu$ are quantities having common values on a universality class. In the $Z_2$ universality class to which the QCD CPs belong, their values are
\begin{align}
  b_4\simeq1.604, \qquad \nu\simeq0.63.
  \label{eq:B4Z2}
\end{align}

In Ref.~\cite{Kiyohara:2021smr}, the behavior of the Binder cumulant around the CP in the heavy-quark region has been investigated in the LAT with large spatial volumes up to the aspect ratio $LT=12$. To realize the simulations with these $LT$, the lattice spacing is fixed to a relatively coarse value $a=1/(4T)$ and the quark determinant is evaluated by the hopping parameter expansion up to the next-to-leading order.

In Fig.~\ref{fig:B4}, we show the result of the Binder cumulant of the Polyakov loop $B_4^\Omega$ for various values of $LT$ as a function of $\lambda\sim\kappa^4$ with $\kappa$ the hopping parameter~\cite{Kiyohara:2021smr}. These numerical results are then fitted by an ansatz inspired by Eq.~(\ref{eq:B4t}). When the results with large spatial volume $LT\ge9$ are used for the fit, it gives $b_4=1.630(24)$ and $\nu=0.614(48)$, which are almost consistent with Eq.~(\ref{eq:B4Z2}) within $1\sigma$. The scaling of the $Z_2$ universality class thus is confirmed with high precision. We also found that the values of $b_4$ and $\nu$ become inconsistent with Eq.~(\ref{eq:B4Z2}) when the results for smaller spatial volumes are included. This result shows that the large spatial volumes are needed to apply the FSS analysis of the Binder cumulant.

%
%

\end{document}